\documentclass[apjl]{emulateapj}

\usepackage{apjfonts}
\usepackage{graphicx,natbib}

\newcommand{\be}{\begin{equation}}
\newcommand{\ee}{\end{equation}}
\newcommand{\ba}{\begin{eqnarray}}
\newcommand{\ea}{\end{eqnarray}}
\newcommand{\Mc}{{\cal M}}
\newcommand{\Ms}{M_{\odot}}
\newcommand{\m}{\langle}
\newcommand{\M}{\rangle}

\newcommand{\bml}{\begin{mathletters}}
\newcommand{\eml}{\end{mathletters}}

\def\ltsima{$\; \buildrel < \over \sim \;$}
\def\simlt{\lower.5ex\hbox{\ltsima}}
\def\gtsima{$\; \buildrel > \over \sim \;$}
\def\simgt{\lower.5ex\hbox{\gtsima}}
\shorttitle{LISA astronomy of double white dwarfs}
\shortauthors{Stroeer et al.}

\begin{document} 

\title{LISA astronomy of double white dwarf binary systems}

\author{A. Stroeer, A. Vecchio}

\affil{School of Physics and Astronomy, 
University of Birmingham, Edgbaston, Birmingham B15 2TT, UK}
\author{G. Nelemans}
\affil{Department of Astrophysics, Radboud University Nijmegen,
Toernooiveld 1, NL-6525 ED Nijmegen, The Netherlands}

\begin{abstract}
The Laser Interferometer Space Antenna (LISA) will provide the largest
observational sample of (interacting) double white dwarf binaries,
whose evolution is driven by radiation reaction and other effects,
such as tides and mass transfer. We show that, depending on the
actual physical parameters of a source, LISA will be able to provide
very different quality of information: for some systems LISA can test
unambiguously the physical processes driving the binary evolution, for
others it can simply {\em detect} a binary without allowing us to
untangle the source parameters and therefore shed light on the physics
at work. We also highlight that simultaneous surveys with GAIA and/or
optical telescopes that are and will become available 
can radically improve the quality
of the information that can be obtained.

\end{abstract}


\maketitle

\section{introduction}
White dwarf (WD) binary systems are the most abundant class of compact
binaries in the Galaxy. Double white dwarf binaries (DWDs) come
in two distinct flavours: detached binaries (WD-WD), that are driven
to shorter and shorter orbital periods primarily by loss of angular
momentum through gravitational wave radiation and interacting white
dwarf binaries (AM CVn systems), where matter is being
transferred from the lower-mass white dwarf (the donor
star) to the more massive accretor. These two classes can be unified
in one picture, in which detached systems evolve to
shorter periods with tidal effects becoming progressively more
important, until mass transfer ensues. Depending on the mass ratio of
the WD-WD, the system will either merge (when the mass ratio is larger
than about 2/3) or reach a ``turning-point'' after which the period
evolution is reversed and the binary evolves to longer periods, with
ever decreasing mass ratio (e.g. Nather et al. 1981, Tutukov \&
Yungelson 1996, Nelemans et al. 2001a, Marsh et al., 2004).

These binaries are important laboratories for some fundamental
questions in astronomy: binary evolution, the structure of WDs, tidal
effects as well as the nature of the progenitors of type Ia supernovae
(e.g. Tout 2005).  Despite their large population, the present sample
of WD binary systems is fairly limited; WDs are electromagnetically
faint and spectroscopic surveys suffer from strong observational
biases, in particular against systems characterised by short orbital
periods, which disappear quickly from the observable sample through
merging.  The ESO SPY survey (Napiwotzki et al. 2004) has
undertaken an extensive spectroscopic survey to search for WD-WD
systems with the primary aim of testing the Supernova Ia
scenario. This has led to the detection of about $\sim 100$ WD-WD
systems, which represents a $\approx 10$-fold increase of the number
of known systems with respect to the pre-SPY era (Marsh
2000). The 17 known AM CVn systems are all found as by-products,
mainly from quasar surveys (see Warner 1995 and Nelemans 2005 for
reviews).

The Laser Interferometer Space Antenna (LISA) (Bender et al, 1998), an
ESA/NASA gravitational wave observatory with launch date 2013+ is
expected to dramatically increase the number statistics of observed
white dwarf binaries by a factor $\simgt 10$ by observing the
gravitational radiation emitted by these systems. LISA will provide
the largest observational sample of these objects, with $10^3 - 10^4$
systems detected during the mission lifetime (e.g. Nelemans et al,
2001b); in addition, such survey is not hampered by the same biases of
those carried out in the electromagnetic band: LISA will probe the
whole galaxy (and possibly beyond to reach the Small and Large
Magellanic Clouds) {\em and} will be very effective in identifying
short period ($\simlt 30$ min) binaries. A handful of interacting
binaries known from previous surveys are so close and with
sufficiently short periods that gravitational waves (GWs) emitted by
them can be confidently detected within a few weeks-to-months of the
operation of the instrument. Indeed DWDs are guaranteed sources
for LISA.  It is also expected that a fair fraction of the LISA
detected systems will be observable with X-ray, optical and possibly
astrometric instruments (such as GAIA), largely increasing the amount
of information that can be extracted (e.g. Nelemans et al. 2004,
Cooray et al, 2004).

Despite the crucial role of DWDs in LISA observations, studies of GW
astronomy have been devoted to either monochromatic detached WD-WD or
WD binaries where the drift in frequency is assumed linear and
``clean'' (e.g. Takahashi and Seto, 2002; Seto 2002), in the sense
that it is exactly described by the evolution of two point-masses in
vacuum. In fact, the astrophysical scenario is radically different: by
the time the frequency evolution becomes important, tidal and other
effects play such an important role that on the one hand the frequency
evolution offers a unique opportunity to studying these effects, and
on the other hand astronomical observations that have been suggested
possible with LISA -- such as the direct measure of the source
distance $D$ and chirp mass $\Mc$ -- will be much more difficult, 
because of the
challenge of disentangling in an unambiguous way several physical
effects.

This raises two key questions that are addressed for the first time in
this Letter: (i) Can LISA test unambiguously the physical process(es)
driving the evolution of a WD binary? As a consequence, (ii) What is
the information that can be gained from the data, if one drops the
{\em a priori} (and unrealistic) assumption that radiation reaction is
the only mechanism at work?  The answers to these questions depend on
a wide range of factors that we discuss, in particular the source
period and mass and whether astrometric and/or optical observations
are available to complement LISA.  

\section{Astronomy with LISA}
\label{sec:astro}

WD binaries enter the LISA observational window (0.1 mHz - 100 mHz)
when they reach a period $P \approx 5$ hr; they secularly evolve
through radiation reaction across the LISA band and are so numerous to
create a stochastic foreground that dominates the LISA observational
window up to $\approx 3$ mHz (e.g. Hils et al, 1990).
When a binary has reached a gravitational wave frequency $f = 2/P
\approx $ a few mHz (and can therefore be resolved individually), the
recorded signal, over a typical observation time $T = 1$ yr, shows an
intrinsic frequency evolution. Depending (primarily) on the mass
ratio, the fate of the binary can be dramatically
different. Comparable mass stars continue their inspiral until they
become unstable and merge in the frequency band 10-100 mHz (depending
on their mass). If the lower mass companion is sufficiently light,
stable mass transfer can start. The orbit eventually stalls --
the frequency evolution as a function of time reaches a ``turning
point'' -- and then the period increases with secular evolution
dominated by mass transfer: the binary becomes observable as an AM CVn
system.

LISA is an all-sky observatory; it monitors DWDs (and other sources) 
providing two independent data sets synthesised
using the technique known as time-delay interferometry (TDI), see
Dhurandhar and Tinto (2005) for a recent review. Each data set (that
we label with $a = 1,2$) can be formally represented as
$s^{(a)}(t) = h^{(a)}(t;\vec{\lambda}) + n^{(a)}(t)$, 
where $h^{(a)}(t;\vec{\lambda})$ is the GW strain, $\vec{\lambda}$ the
vector of the unknown signal parameters and
$n^{(a)}(t)$ the noise. Here we model the gravitational waveform at  
the lowest Newtonian quadrupole order -- 
in fact post-Newtonian corrections to the phase
contribute $\ll 1$ wave cycle for multi-year observations and
harmonics at other multiples of $1/P$ are a factor
$\simlt 10^3$ smaller than the leading order 
Newtonian quadrupole (Blanchet et al,
1996) -- and the noise according to the analytical fit by Barack and Cutler 
(2004), ignoring the possible contribution from
unresolved extreme mass ratio in-spirals. We 
 model the detector output using the so-called
rigid adiabatic approximation (Rubbo et al, 2004), so that the
signal registered at LISA outputs can 
be represented as (Vecchio \& Wickham, 2004)
\be
h^{(a)}(t) = A_0\,\sum_{n = 1}^4 A_n^{(a)}(t)\,\cos\chi_n^{(a)}(t)\,,
\label{hl1}
\ee
where 
\be
A_0 = 2\,(\pi f_0)^{2/3}\,\frac{\Mc^{5/3}}{D}\,;
\label{f:A}
\ee
$A_n^{(a)}[t; {\bf \hat N}\,,{\bf \hat L}, f(t)]$ and 
$\cos\chi_n^{(a)}[t;{\bf \hat N}\,,{\bf \hat L},f(t)]$ 
contain information about the intrinsic frequency
evolution $f(t)$ -- here $f_0$ is the frequency at the beginning of 
the observations --  and the source position in the sky ${\bf \hat N}$ and the
orientation of its orbital plane ${\bf \hat L}$; explicit expressions 
for $h^{(a)}(t)$ can be found {\em e.g.} in Vecchio \& Wickham (2004).

The intrinsic frequency of radiation from a DWD slowly changes 
during the typical LISA observation time. From a phenomenological point of
view, we can therefore Taylor expand it around an arbitrary fixed time 
(say the beginning of the observations $t_0 = 0 $), 
so that the phase of the signal can be written as:
\be
\phi(t) = \phi_0 + 2 \pi f_0 t + 2 \pi \sum_{k = 1}^K
\frac{t^{(k+1)}}{(k+1)!}\,\frac{d^kf}{dt^k}\,,
\label{e:fTaylor}
\ee 
where we define with $f^{(k)} \equiv {d^kf}/{dt^k}$ ($k = 1,...,K$) the
``spin-down parameters'' -- notice that despite the terminology $f^{(k)}$ could
be either positive or negative. The number of spin-down parameters that
is retained into the phase model depends on the actual source parameters and is
determined by the condition that the integrated value of the phase over
the whole observation time differs from the actual one by $\ll 1$. 
If both $\dot{f}_0$ and
$\ddot{f}_0$ are observable ($K = 2$), then one can measure the so-called braking 
index $n \equiv {f_0\,\ddot{f}_0}/{\dot{f}_0^2}$, which provides a powerful
tool to discriminate the physical mechanism driving the orbital evolution;
if only radiation reaction is at work, then $n = 11/3$.

The signal $h^{(a)}(t;\vec{\lambda})$ depends therefore on a vector $\vec{\lambda}$ of
$7+K$ unknown parameters: the amplitude $A_0$, four angles that identify
${\bf \hat N}$ and ${\bf \hat L}$, the frequency at the
beginning of the observation $f_0$, an overall arbitrary phase $\phi_0$ and 
$K$ spin-down parameters. We can then estimate the expected
mean-square-error $\m (\Delta\lambda^j)^2\M$ ($j$ labels the components of 
$\vec{\lambda}$) which is associated to the LISA measurement of each parameter
$\lambda^j$ by computing the variance-covariance matrix (Vecchio \&
Wickham, 2004); the diagonal elements of the above matrix 
provide (in the limit of high
signal-to-noise ratio $\rho$) a tight 
{\em lower limit} to $\m (\Delta\lambda^j)^2\M$ 
(see, {\em e.g.} Nicholson and Vecchio, 1998).

A thorough exploration of the parameter space characterising DWDs goes
beyond the scope of this paper. Here we concentrate on two
fiducial sources that are representative of the astrophysical scenarios
considered here (Marsh et al, 2004): (i) An ``unstable'' binary with
$m_1 = m_2 = 0.6\,\Ms$, that never reaches a turning point and merges
at 37.7 mHz; in order to faithfully describe the phase evolution one
needs to include $\dot{f}_0$ for $f_0 \ge 0.3$ mHz (1 mHz) and both
$\dot{f}_0$ and $\ddot{f}_0$ for $f \ge 8.8$ mHz (21.7 mHz) for an
integration time $T = 5$ yr (1 yr); (ii) A ``stable'' binary with
individual progenitor masses $m_1 = 0.4\,\Ms$ and $m_2 = 0.2\,\Ms$,
whose frequency increases until it reaches a turning point at $f =
9.7$ mHz and then enters the AM CVn phase where the frequency {\em
decreases}; the phase evolution is described by only the first
spin-down parameter $\dot{f}_0$ in the frequency range
$0.5\,\mathrm{mHz} \,(1.2\,\mathrm{mHz})\le f_0 \le 9.7$ mHz (in the
inspiral phase) and $0.8\,(2.6\,\mathrm{mHz}) \le f_0 \le 9.7$ mHz (in
the AM CVn phase) for $T = 5$ yr (1 yr); elsewhere
the source is detected as monochromatic. In our calculations we take 
tidal effects and mass transfer explicitly into
account. However, we do not consider short timescale variations that are
observed in many binaries and which might have a (strong) influence on 
the instantaneous value of spin-down parameters (e.g. Marsh \& Nelemans, 2005).


\begin{table}[t!]
\caption{\label{tab:res} 
Signal-to-noise ratio $\rho$ and the statistical errors associated to
the measurement of the amplitude $A$, the position of the source in
the sky $\Delta \Omega_N$, the inclination $\cos\iota$, the
spin-down parameters, $\dot{f}_0$ and $\ddot{f}_0$, and the braking
index $n$ in observation of an unstable binary at 30 mHz (top) and 20
mHz (middle) and a stable AM CVn system at 8 mHz (bottom). We
list the median and the value corresponding to a cumulative
probability of 90\% of the separate distributions for each
parameter. In each case we simulate $10^4$ sources at distance $D =
10$ kpc with random sky position and orientation of the orbital plane. 
The observation time is $T = 1$ yr and $5$ yr.}
\begin{center}
 \begin{tabular}{|l|c|c|c|c|c|c|c|c|}
\hline
 & T & $\rho$ & ${\Delta A}/{A}$ & $\Delta\Omega_N$ & $\Delta\cos\iota$ & ${\Delta\dot{f}_0}/{\dot{f}_0}$ & ${\Delta\ddot{f}_0}/{\ddot{f}_0}$ & $\Delta n$ \\
& (yr) &     &              &     (deg$^2$)       &      &                            &  & \\
\hline
 median &1    &  23 &  0.13&  0.09 &   0.31&    $1.3\times 10^{-3}$ &  -- &  -- \\
 {90\%} &1    &  14 &  0.30&  0.29 &   0.68&    $2.3\times 10^{-3}$ &  -- &  -- \\
 median &5    &  52 &  0.06&  0.004&   0.2&     $4.1\times 10^{-5}$ &  $8.1\times 10^{-3}$ &  0.03\\
 90\%   &5    &  32 &  0.14&  0.01 &   0.4&     $6.5\times 10^{-5}$ &  $1.3\times 10^{-2}$ &  0.05\\
 \hline
 median &1    &  33 &  0.09&  0.12 &   0.26&    $4.4\times 10^{-3}$ &  -- &  --\\
 90\%   &1    &  20 &  0.20&  0.31 &   0.56&    $7.4\times 10^{-3}$ &  -- &  --\\
 median &5    &  74 &  0.04&  0.005&   0.17&    $1.3\times 10^{-3}$ &  $7.6\times 10^{-2}$ &  0.28\\
 90\%   &5    &  45 &  0.09&  0.01 &   0.37&    $2.1\times 10^{-3}$ &  $1.2\times 10^{-1}$ &  0.45\\
 \hline
 median &1    &  8.1&  0.35&  11.75&   0.52&    1.6                 &  --&  -- \\
 90\%   &1    &  5.0 &  0.82&  30.93&  1.10 &    2.8                 &  --&  -- \\
 median &5    &  18 &  0.15&  0.43 &  0.34 &    $8.0\times 10^{-3}$ &  --&  -- \\
 90\%   &5    &  11 &  0.40&  1.08 &  0.74 &    $1.3\times 10^{-2}$ &  --&  -- \\
 \hline
 \end{tabular}
\end{center}
 \end{table}

In order to quantify how accurately LISA can measure the
parameters of interacting WDs, we have carried out Monte
Carlo simulations over a population of $10^4$ sources characterised by
the same physical parameters -- $A_0$, $f_0$, $\dot{f}_0$ and
$\ddot{f}_0$, if relevant -- but with random values of ${\bf \hat N}$
and ${\bf \hat L}$. All the sources are placed at the same
distance $D =10$ kpc (which in turn sets $A_0$). We simulate a
population of unstable binaries at $f_0 = 20$ mHz and $f_0 = 30$ mHz,
respectively, where the orbital separation is $\approx 5$ star radii.
These frequencies -- chosen to provide a ``best case scenario'' from the
point of view of information extraction --  
are at the very high end of the range expected for Galactic DWDs
(Nelemans et al, 2001b). We also simulate 
a population of stable binaries at $f_0 = 8$ mHz in
the AM CVn phase, where $m_1 = 0.43\,\Ms$, $m_2 = 0.17\,\Ms$ and 
$\dot{f}_0 = - 1.2 \times 10^{-15}\,\mathrm{s}^{-2}$. We compute the
variance-covariance matrix and summarise the results of the
distributions of the expected mean-square-errors associated to the
LISA measurements in Table 1. 

We consider now what are the
consequences for astronomy and concentrate on the specific results for
$T = 5$ yr; the reader can easily re-scale the conclusion for $T = 1$
yr, using the results provided in Table 1. 
About a
third of the DWDs detectable by LISA are also expected to be observed
as eclipsing binaries (Cooray et al, 2004) yielding a direct measure
of $R_1/(a\,\cos\iota)$, $R_2/(a\,\cos\iota)$ and $R_1/R_2$, where
$R_{1,2}$ are the radii of the two stars, $a$ is the orbital
separation and $\cos\iota = {\bf \hat N} \cdot {\bf \hat L}$ the orbital
inclination. In addition, about half of the DWDs might be detected
by GAIA; according to Cooray et al. (2004) the distribution of the
V-magnitude of galactic DWDs peak around 20  which corresponds to the
magnitude limit up to which GAIA can do high precision 
observations\footnote{However, the estimate by Cooray et al. (2004)
is based on very optimistic
calculations of tidal heating by Iben et al. (1998)}.
The astrometric accuracy of the instrument is $200\,\mu\mathrm{as}$
for $V = 20$, $10\,\mu\mathrm{as}$ for $V = 15$, and
$4\,\mu\mathrm{as}$ for $V < 12$ which implies that GAIA can detect
DWDs up to a few kpc and measure their distance with an error in the
range $\approx 1\% - 10\%$. The possibility of observing DWDs with
LISA, GAIA and/or spectroscopically may be crucial in disclosing the
physics at work in these systems by breaking the degeneracy of a large
number of parameters, as we now discuss in more detail.

We focus first on the unstable DWD for which both $\dot{f}_0$ and
$\ddot{f}_0$ are observable. This is crucial, because one can measure
the braking index with an absolute error $\Delta n \approx 0.1$, and
LISA will be able to test directly the physical processes that
drive the binary evolution. If observations are consistent with the
assumption that the evolution is driven by radiation reaction only,
then from $\dot{f}_0 = ({96}/{5})\,\pi^{8/3}\,\Mc^{5/3}\,f^{11/3}$ one
can measure directly the chirp mass of the binary with a statistical
error $\m (\Delta \Mc/\Mc)^2 \M^{1/2} \simeq (3/5)\,\m (\Delta
\dot{f}_0/\dot{f}_0)^2 \M^{1/2}\sim 10^{-4}$ and, through
Eq.~(\ref{f:A}) the distance with $\m (\Delta D/D)^2 \M^{1/2} \simeq
\m (\Delta A/A)^2 \M^{1/2}\approx 0.1$ (in both cases we have ignored
the errors on $f_0$ and the correlations that are negligible).  If a
LISA source is also observed as an eclipsing binary, then one is able
to measure $R_1$ and $R_2$ ($\cos\iota$ is estimated with a good
accuracy $\approx 10\%$ by LISA) and place an upper-limit on the total
mass of the system: these combined observations represent a unique
opportunity to disclose the physics of compact objects.

However, our theoretical understanding of DWDs, despite limited,
suggests that by the time a source emits in the mHz region the
orbital evolution is not clean, which, for the particular source we
are considering now, will be unveiled by LISA through a direct measure
of $n$ not consistent with $11/3$. In this case $\dot{f}_0$ and
$\ddot{f}_0$ carry the signature not only of $\Mc$ but also of the
parameters that control tidal effects.  If only LISA data are
available, then neither $\Mc$ nor $D$ can be inferred (only loose
limits can be placed): this is in strong contrast with what has been
assumed so far, based on the naive hypothesis that hydrodynamical
effects can be neglected.  Nonetheless, the situation is radically
different if the LISA source is also observed by GAIA which provides
an independent measure of $D$: from it and the amplitude of the
gravitational signal, Eq.~(\ref{f:A}), one can estimate $\Mc$ with an
error $\m \left(\Delta \Mc/\Mc\right)^2 \M^{1/2} = (3/5) \sqrt{\m
\left(\Delta D/D\right)^2\M + \m \left(\Delta A/A\right)^2\M} \approx
0.1$, if one assumes that $\m (\Delta D/D)^2\M^{1/2} \approx \m
(\Delta A/A)^2\M^{1/2} \approx 0.1$; for some systems one could
achieve an accuracy as good as $\approx 0.01$.  In other words, GAIA
re-instates the full power of LISA observations by breaking the
degeneracy between $\Mc$ and $D$. From $\Mc$, $\dot{f}_0$ and
$\ddot{f}_0$ one can learn new physics that we have not been able to
explore directly so far. Of course the most exciting situation is if
the LISA and GAIA source is also identified optically as an eclipsing
binary, which provides additional information on the stars size and
total mass.

We turn now attention on the observations of the stable AM CVn
progenitor. The key difference from the scenario described above is
that LISA is now able to detect only $\dot{f}_0$; moreover the
signal-to-noise is much smaller, which degrades the quality of
astronomy in general, and the size of the LISA error box in the sky in
particular: the association of a source observed by LISA with stars
detected by other surveys becomes much more challenging. If LISA measures a
value $\dot{f}_0 < 0$, then this provides a strong evidence that
the DWD evolution is driven by mass transfer; in this case however,
$D$ can not be inferred because $\dot{f}_0$ depends on $\Mc$ and other
physical parameters; the chirp mass is equally unmeasurable. On the
other hand, if $\dot{f}_0 > 0$, then one is not able to make any
statement about whether the system is ``clean'' or not. An independent
determination of the distance ({\em e.g.} through GAIA) becomes again
essential: if this is possible, one can compare the value of
$\dot{f}_0$ predicted by evolution driven purely by radiation reaction
with the value obtained with LISA. For AM CVn the ability of testing
in detail models of orbital evolution and mass transfer is somewhat
reduced because only one parameter ($\dot{f}_0$) is directly measured;
from this point of view, a mission lasting up to 10 years (which is
not a mission requirement at present but is considered feasible with
the instrumentation on board) is crucial, as $\ddot{f}_0$ becomes
observable for an increasingly larger number of sources. It is also worth
emphasising that if an AM CVn is observed optically, then one might
measure the {\em super-hump} period. Based on extrapolation of an
empirical relation, this yields the mass
ratio of the binary with an accuracy of a few tens of percent
(e.g. Espaillat et al. 2005).  From $\Mc$ (provided by LISA {\em and}
GAIA) and $m_1/m_2$, one can determine the individual masses of the
stars and, combined with $\dot{f}_0$, the mass transfer rate.

\section{Conclusion}

Interacting WD binaries are unique laboratories for testing
theories about binary evolution, the structure of white dwarfs, tidal
effects, mass transfer and supernova-Ia progenitors. LISA opens a
radically new opportunity to carry out these studies by providing a
much larger (and unbiased) sample of these stars and information on
physical parameters that are hardly accessible by other
surveys. However the challenge that LISA astronomers will face is to
untangle a large number of physical parameters from a few
phenomenological observables.  

We have shown that if the first two
time derivatives of the frequency, $\dot{f}_0$ and $\ddot{f}_0$, are
measured, then LISA alone will be able to test accurately whether or
not a system evolves cleanly due to radiation reaction. If the
evolution is driven by the competition of several processes (which we
expect in a large number of cases), then the phase
evolution monitored by the instrument allows us to constrain models of
evolution of WD binaries. If only $\dot{f}_0$ is observed, then LISA
provides no information on the problem at hand. If the evolution is
not ``clean'' and only LISA data are available, one is not able to
infer the chirp mass and distance of the binary: this is in
strong contrast with what has been assumed so far based on the
simplistic assumption that tidal effects and mass transfer can be
ignored.  However, if a LISA source is also observed by GAIA the
situation changes dramatically: the combined observations provide both
$\Mc$ and $D$ with an accuracy $\approx 10\%$. Moreover combining LISA
and GAIA observations, it would still be possible to test whether the
binary evolution is clean even if LISA can measure only
$\dot{f}_0$. In addition, a fair fraction of DWDs detected with LISA
are observable as eclipsing binaries, from
which one can determine the stars individual radii: this opens the
possibility of extracting an unprecedented wealth of information about
the physics of these compact objects in the most extreme physical
conditions. 

Our study calls for a more detailed investigation of
this observational scenario over the full parameter space (such work
is presently in progress and will be reported elsewhere) including
the exploitation of the full power of the TDI observables 
(e.g. Dhurandhar et al, 2002, Prince et al, 2002) and detailed
theoretical modelling of interacting white dwarfs in realistic
conditions in order to be able to link reliably the phenomenological
observational parameters to those that underpin the key physics.

\acknowledgments
GN is supported by NWO-VENI grant 639.041.405.

\end{document}